\documentclass[epj]{svjour}

\usepackage[dvips]{graphicx}

\usepackage{array}
\usepackage{amssymb}
\usepackage{amsmath}
\usepackage{hhline}
\usepackage{longtable}
\usepackage{dcolumn}
\usepackage{bm}
\usepackage{subfigure}
\usepackage{epsfig}
\usepackage{latexsym,amsmath}
\usepackage{amsbsy}

\begin{document}
\title{Finite epidemic thresholds in fractal scale-free `large-world' networks}

\author{Zhongzhi Zhang\inst{1,2} \thanks{e-mail: zhangzz@fudan.edu.cn} \and Shuigeng Zhou\inst{1,2} \thanks{e-mail: sgzhou@fudan.edu.cn} \and Tao Zou\inst{1,2}  \and Jihong Guan\inst{3}}                     
\institute{Department of Computer Science and Engineering, Fudan
University, Shanghai 200433, China \and Shanghai Key Lab of
Intelligent Information Processing, Fudan University, Shanghai
200433, China \and Department of Computer Science and Technology,
Tongji University, 4800 Cao'an Road, Shanghai 201804, China}

\date{Received: date / Revised version: date}

\abstract{It is generally accepted that scale-free networks is prone
to epidemic spreading allowing the onset of large epidemics whatever
the spreading rate of the infection. In the paper, we show that
disease propagation may be suppressed in particular fractal
scale-free networks. We first study analytically the topological
characteristics of a network model and show that it is
simultaneously scale-free, highly clustered, ``large-world", fractal
and disassortative. Any previous model does not have all the
properties as the one under consideration. Then, by using the
renormalization group technique we analyze the dynamic
susceptible-infected-removed (SIR) model for spreading of
infections. Interestingly, we find the existence of an epidemic
threshold, as compared to the usual epidemic behavior without a
finite threshold in uncorrelated scale-free networks. This
phenomenon indicates that degree distribution of scale-free networks
does not suffice to characterize the epidemic dynamics on top of
them. Our results may shed light in the understanding of the
epidemics and other spreading phenomena on real-life networks with
similar structural features as the considered model.
} 

 \maketitle
\section{Introduction}

In the past ten years, there  has been a considerable interest in
characterizing and understanding the topological properties of
networked systems~\cite{AlBa02,DoMe02,Ne03,BoLaMoChHw06,CoRoTrVi07}.
It has been established that scale-free behavior~\cite{BaAl99} is
one of the most fundamental concepts constituting our basic
understanding of the organization of many real-world systems in
nature and society. This scale-free property has a profound effect
on almost every aspect on dynamic processes taking place on
networks, including robustness~\cite{AlJeBa00},
percolation~\cite{CaNeStWa00,CoErAvHa01},
synchronization~\cite{WaCh02}, games~\cite{SzFa07}, epidemic
spreading~\cite{PaVe01a,PaVe01b,MoPaVe02}, and so on. For instance,
for a wide range of scale-free networks, there is no existence of an
epidemic threshold, even infections with low spreading rate will
prevail over the entire population in these
networks~\cite{PaVe01a,PaVe01b,MoPaVe02}. This radically changes the
conclusion drawn from classic disease modeling~\cite{He00}.

Recently, it has been discovered that many real-life networks, such
as the WWW, metabolic networks, and yeast protein interaction
networks have self-similar properties and exhibit fractal
scaling~\cite{SoHaMa05,SoHaMa06,YoRaOr06,GoSaKaKi06}. The fractal
topology can be characterized through two exponents: fractal
dimension $d_B$ and degree exponent of the boxes $d_k$, which can be
obtained by box-counting
algorithm~\cite{SoHaMa05,SoHaMa06,SoGaHaMa07,KiGoKaKi07}. The
scaling of the minimum number of boxes $N_{B}$ of linear size
$\ell_{B}$ needed to cover the network with node number $N$ defines
the fractal dimension $d_B$, namely $N_{B}/N\thicksim
\ell_{B}^{-d_{B}}$. Analogously, the degree exponent of the boxes
$d_k$ is identified through $k_{B}(\ell_{B})/ k_{hub} \thicksim
\ell_{B}^{-d_{k}}$, where $k_{B}(\ell_{B})$ is the number of
outgoing links from the box as a whole, and $k_{hub}$ the largest
node degree inside the box. Fractal networks are all self-similar,
which means that fractal scale-free networks present the property of
scale-invariance of degree distribution, $P(k)\sim k^{-\gamma}$,
i.e., the exponent $\gamma$ remains the same for different box
sizes~\cite{SoHaMa05}. In self-similar scale-free networks, the
three indexes $\gamma$, $d_{B}$ and $d_{k}$ satisfy the following
relation: $\gamma =1+d_{B}/ d_{k}$~\cite{SoHaMa05}.

As a fundamental property, topological fractality relates to many
respects of network structure and function. Recent authors have
shown that the correlation between degree and betweenness centrality
of nodes is much weaker in fractal network models in comparison with
non-fractal models~\cite{KiHaPaRiPaSt07}. It has been also
shown~\cite{SoHaMa06,YoRaOr06,ZhZhZo07} that fractal scale-free
networks are not assortative, this disassortativity feature together
with fractality makes such scale-free networks more robust against
intentional attacks on hub nodes, as compared to the very vulnerable
non-fractal scale-free networks~\cite{SoHaMa06}. In addition to the
distinction in the robustness, fractal networks exhibit poorer
synchronizability than non-fractal counterparts~\cite{ZhZhZo07}.
Although a lot of jobs have been devoted to characterizing fractal
networks~\cite{BaFeDa06,GaSoHaMa07,CoBeTeVoKl07,Hi07,RoHaAv07,RoAv07},
it is still of current interest to model fractal topology and seek a
better understanding of its consequences on dynamic processes.

In this paper, we relate fractality to dynamics of disease spread in
deterministic networks. Deterministic graphs have strong advantages.
For example, they allow to compute analytically their properties,
which have played a significant role, both in terms of explicit
results and a guide to and a test of simulated and approximate
methods~\cite{BaRaVi01,DoGoMe02,CoFeRa04,ZhRoZh07,ZhZhCh07,JuKiKa02,ZhZhChGuFaZa07,RaSoMoOlBa02,RaBa03,ZhZhFaGuZh07,AnHeAnSi05,DoMa05,ZhCoFeRo05,ZhRo05,CoOzPe00,CoSa02,ZhRoGo05,ZhWaHuCh06,ChYuXuCh07,ZhZhWaSh07}.
We first introduce a deterministic family of fractal graphs. From
the viewpoint of complex networks, we determine accurately the
topological characteristics of a particular graph and show that it
is simultaneously scale-free, highly clustered, fractal and
disassortative, but lacks the small-world property. Then we define a
dynamic susceptible-infected-removed (SIR) model~\cite{He00} on the
two dimensional fractal graph to study the effect of fractality on
disease spreading. By mapping the SIR model to a bond percolation
problem, we found that there is an existence of finite epidemic
threshold. Thus, the transmission rate needs to exceed a critical
value for the disease to spread and prevail, which shows that the
fractal networks are robust to infection.

\section{Network construction and topologies}

This section is devoted to the construction and the relevant
structural properties of the studied network, such as degree
distribution, clustering coefficient, average path length (APL),
fractality, and correlations.

\begin{figure}[h]
\centering\includegraphics[width=0.45\textwidth]{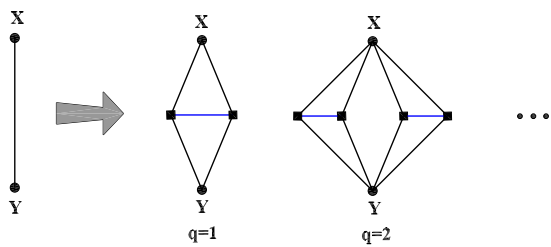}
\caption{(Color online) Iterative construction method of the fractal
networks. Each iterative link is replaced by a connected cluster on
the right-hand side of the arrow. The blue links are noniterated
ones.}\label{fig1}
\end{figure}

\subsection{Construction algorithm}
\begin{figure}
\includegraphics[width=0.5\textwidth]{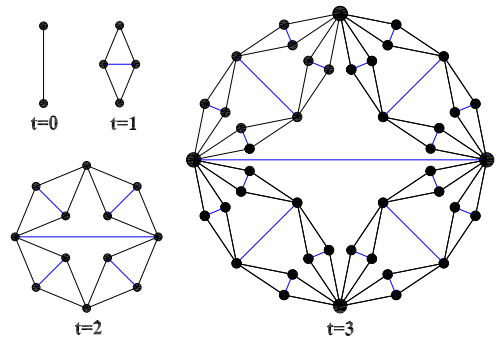}
\caption{(Color online) Scheme of the growth of the network for the
particular case of $q=1$. Only the first three iterative processes
are shown.}\label{fig2}
\end{figure}

The proposed class of fractal networks is constructed in an
iterative way as shown in Fig.~\ref{fig1}. Let $F_{t,q}$ ($t\geq 0$,
$q\geq 1$) denote the networks after $t$ iterations. Then the
networks are generated as follows: For $t=0$, $F_{0,q}$ is an
iterative edge connecting two nodes. For $t\geq 1$, $F_{t,q}$ is
obtained from $F_{t-1,q}$. We replace each existing iterative edge
in $F_{t-1,q}$ by a connected cluster of edges on the right of
Fig.~\ref{fig1}, where $q$ denotes the number of the newly-created
noniterated links induced by the iterative edge. The growing process
is repeated $t$ times, with the family of fractal graphs obtained in
the limit $t \to \infty$. Figure~\ref{fig2} shows the growing
process of the network for the particular case of $q=1$. Note that
all generated networks have qualitatively similar properties. In
what follows we focus on the special case of $q=1$ and denote it as
$F_{t}$ after $t$ iterations.

Next we compute the numbers of total nodes (vertices) and links
(edges) in $F_t$. Notice that there are two types of links (i.e.,
iterative links and noniterated links) in the network. Let $L_v(t)$,
$L_i(t)$ and $L_n(t)$ be the number of new vertices, iterative
links, and noniterated links created at step $t$, respectively.
Since all old iterative links are not preserved in the growing
process, thus $L_i(t)$ is in fact the total number of iterative
links at time $t$. Note that each of the existing iterative links
yields two nodes connected by one noniterated link, and the addition
of each new node leads to two iterative links. By construction, for
$t\geq 1$, we have
\begin{equation}
\left\{\begin{array}{lc} {\displaystyle{L_i(t)=4\,L_i(t-1),} }\\
{\displaystyle{L_v(t)=2\,L_i(t-1),} }\\
{\displaystyle{L_n(t)=L_i(t-1).} }\end{array} \right.
\end{equation}

Considering the initial condition $L_v(0)=2$, $L_i(0)=1$, and
$L_n(0)=0$, it follows that
\begin{equation}
\left\{\begin{array}{lc} {\displaystyle{L_v(t)=2\cdot 4^{t-1},} }\\
{\displaystyle{L_i(t)=4^t,} }\\
{\displaystyle{L_n(t)= 4^{t-1}.} }\end{array} \right.
\end{equation}

Thus the number of total nodes $N_t$ and edges $E_t$ present at step
$t$ is
\begin{eqnarray}\label{Nt1}
N_t=\sum_{t_i=0}^{t}L_v(t_i)=\frac{2\cdot4^{t}+4}{3}
\end{eqnarray}
and
\begin{eqnarray}\label{Et1}
E_t=L_i(t)+\sum_{t_i=1}^{t}L_n(t_i)=\frac{4^{t+1}-1}{3},
\end{eqnarray}
respectively. The average degree after $t$ iterations is
\begin{equation}
\langle k \rangle
_t=\frac{2\,E_t}{N_t}=\frac{2(4^{t+1}-1)}{2\cdot4^{t}+4},
\end{equation}
which approaches 4 in the infinite $t$ limit.

\subsection{Degree distribution}

When a new node $u$ is added to the network at step $t_u$ ($t_u\geq
1$), it has three links, among which two are iterative links and one
is noniterated link. Let $L_i(u,t)$ be the number of iterative links
at step $t$ that will create new nodes connected to the node $u$ at
step $t+1$. Then at step $t_u$, $L_i(u, t_u)=2$. From the iterative
generation process of the network, one can see that at any
subsequent step each iterative link of $u$ is broken and generates
two new iterative links connected to $u$. We define $k_u(t)$ as the
degree of node $u$ at time $t$, then the relation between $k_u(t)$
and $L_i(u,t)$ satisfies:
\begin{equation}\label{deltak}
k_u(t)=L_i(u,t)+1,
\end{equation}
where the last term 1 represents the only noniterated link of node
$u$. Now we compute $L_i(u,t)$. By construction,
$L_i(u,t)=2\,L_i(u,t-1)$. Considering the initial condition $L_i(u,
t_u)=2$, we can derive $L_i(u,t)=2^{t-t_{u}+1}$. Then at time $t$,
the degree of vertex $u$ becomes
\begin{equation}\label{ki}
k_u(t)=2^{t-t_{u}+1}+1.
\end{equation}

It should be mentioned that the initial two nodes created at step 0
have a little different evolution process from other ones. Since the
initial two nodes have no noniterated link, we can easily obtain
that at step $t$, for either of the initial two nodes, its degree
just equals the number of iterative links connecting it, both of
which are $2^{t}$.

Equation~(\ref{ki}) shows that the degree spectrum of the network is
discrete.  It follows that the cumulative degree
distribution~\cite{Ne03} is given by
\begin{equation}\label{pcumk}
P_{\rm cum}(k)=\sum_{\tau \leq t_u}\frac{L_v(\tau)}{N_t}
={2\cdot4^{t_u}+4\over 2\cdot4^{t}+4}.
\end{equation}
Substituting for $t_u$ in this expression using
$t_u=t+1-\frac{\ln(k-1)}{\ln 2}$ gives
\begin{equation}
P_{\rm cum}(k)=\frac{2\cdot4^{t}\cdot
4(k-1)^{-(\ln4/\ln2)}+4}{2\cdot4^{t}+4}.
\end{equation}
When $t$ is large enough, one can obtain
\begin{equation}\label{gammak}
P_{\rm cum}(k)=4\,(k-1)^{-2}.
\end{equation}
So the degree distribution follows a power law form with the
exponent $\gamma=3$. The same degree exponent has been obtained in
the famous Barab\'asi-Albert (BA) model~\cite{BaAl99} as well as
some other deterministic
models~\cite{Hi07,ZhZhZo07,RoHaAv07,RoAv07,BeOs79,KaGr81,HiBe06,Ya88}.

\subsection{Clustering Coefficient}

The clustering coefficient~\cite{WaSt98} of a node $u$ with degree
$k_u$ is given by $C_u =2e_u/[k_u( k_u-1)]$, where $e_u$ is the
number of existing links among the $k_u$ neighbors. Using the
construction rules, it is straightforward to calculate analytically
the clustering coefficient $C(k)$ for a single node with degree $k$.
For the initial two nodes born at step 0, their degree is $k=2^t$,
and the  existing links among these neighbors is $\frac{k}{2}$, all
of which are noniterated links. For those nodes created at step
$\phi$ $(0<\phi<t)$, there are only $\frac{k-1}{2}$ links that
actually exist among the neighbor nodes. Finally, for the smallest
nodes created at step $t$, each has a degree of $k=3$, the existing
number of links between the neighbors of each is 2. Thus, there is a
one-to-one correspondence between the clustering coefficient $C(k)$
of the node and its degree $k$:
\begin{equation}\label{Ck}
C(k)=\left\{\begin{array}{lc} {\displaystyle{ 1 / (k-1) } }
& \ \hbox{for}\ k=2^t\\
{\displaystyle{1 / k} }
& \ \hbox{for}\  k=2^m+1 (2 \leq m\leq t)\\
{\displaystyle{ 2 / k } } & \ \hbox{for} \  k=2^1+1\end{array}
\right.
\end{equation}
which is inversely proportional to $k$ in the limit of large $k$.
The scaling of $C(k)\sim k^{-1}$ has been observed in many
real-world scale-free networks~\cite{RaBa03}.

Using Eq.~(\ref{Ck}), we can obtain the clustering $C_t$ of whole
the network at step $t$, which is defined as the average clustering
coefficient of all individual nodes. Then we have
\begin{equation}\label{ACC}
C_t=
    \frac{1}{N_{t}}\left [\frac{L_v(0)}{D_0-1}+\sum_{r=1}^{t-1}\frac{L_v(r)}{D_r}+ \frac{2\,L_v(t)}{D_t} \right],
\end{equation}
where $D_r$ is the degree of a node at time $t$, which was created
at step $r$, see Eq.~(\ref{ki}). In the infinite network order limit
($N_{t}\rightarrow \infty$), Eq.~(\ref{ACC}) converges to a nonzero
value $\bar{C}=0.5435$. Therefore, the average clustering
coefficient of the network is very high.

\subsection{Fractal dimension}

As a matter of fact, the fractal graph grows as a inverse
renormalization procedure, see Fig.~\ref{fig2} in reverse order. To
find the fractal dimension, we follow the mathematical framework
presented in Ref.~\cite{SoHaMa06}. By construction, in the infinite
$t$ limit, the different quantities grow as:
\begin{equation}\label{Frac01}
\left\{\begin{array}{lc} {N_t\simeq 4\,N_{t-1} },\\
{k_u(t)\simeq 2\,k_u(t-1)},\\
{\mathbb{D}_t=2\,\mathbb{D}_{(t-1)},}\end{array} \right.
\end{equation}
where the third equation describes the change of the diameter
$\mathbb{D}_t$ of the graph $F_t$, where $\mathbb{D}_t$ is defined
as the longest shortest path between all pairs of nodes in $F_t$.

From the relations provided by Eq.~(\ref{Frac01}), it is clear that
the quantities $N_t$, $k_u(t)$ and $\mathbb{D}_t$ increase by a
factor of $f_N=4$, $f_k=2$ and $f_{\mathbb{D}}=2$, respectively.
Then between any two times $t_{1}$, $t_{2}$ ($t_{1} < t_{2}$), we
can easily obtain the following relation:
\begin{equation}\label{Frac02}
\left\{\begin{array}{lc} {\mathbb{D}_{t_{2}}=2^{t_{2}-t_{1}}\,\mathbb{D}_{t_{1}}},\\
{N_{t_{2}}=4^{t_{2}-t_{1}}\,N_{t_{1}}},\\
{k_u(t_{2})=2^{t_{2}-t_{1}}\,k_u(t_{1})}.\end{array} \right.
\end{equation}
From Eq. (\ref{Frac02}), we can derive the scaling exponents in
terms of the microscopic parameters: the fractal dimension is $d_B
=\frac{\ln f_N}{\ln f_{\mathbb{D}}}=2$, and the degree exponent of
boxes is $d_k =\frac{\ln f_k}{\ln f_{\mathbb{D}}}=1$. The exponent
of the degree distribution satisfies $\gamma =1+\frac{d_B}{d_k}=3$,
giving the same $\gamma$ as that obtained in the direct calculation
of the degree distribution, see Eq.~(\ref{gammak}).

Note that in a class of deterministic models called pseudo-fractals,
although the number of their nodes increases exponentially, the
additive growth of the diameter with time implies that the networks
are small world. These models do not capture the fractal topology
found in diverse complex
networks~\cite{DoGoMe02,CoFeRa04,ZhRoZh07,ZhZhCh07,JuKiKa02,ZhZhChGuFaZa07,ZhZhFaGuZh07,AnHeAnSi05,DoMa05,ZhCoFeRo05}.

\subsection{Degree correlation}

Degree
correlation~\cite{MsSn02,PaVaVe01,VapaVe02,Newman02,Newman03c,BoPa03,BoPa05,ZhZh07}
is a particularly interesting subject in the field of complex
networks, because it can give rise to some interesting network
structure effects. Degree correlation in a network can be measured
by means of the quantity, called \emph{average nearest-neighbor
degree} (ANND) and denoted as $k_{nn}(k)$, which is a function of
node degree, and is more convenient and practical in characterizing
degree correlation. ANND is defined by~\cite{PaVaVe01}
\begin{equation}\label{knn01}
  k_{nn}(k) = \sum_{k'} k' P(k'|k),
\end{equation}
where $P(k'|k)$ is the probability that a link from a node of degree
$k$ points to a node of degree $k'$.

For the fractal graph considered here, one can exactly calculate
$k_{\rm nn}(k)$ . By construction, all neighbors of the initial two
nodes have the same degree 3, while for each other nodes with degree
greater than 3, only one of its neighbor has the same degree as
itself, all the rest neighbors have a degree 3. Then we have
\begin{equation}\label{knn02}
\left\{\begin{array}{lc} {\displaystyle{ k_{\rm nn}(k)=3 } }
& \ \hbox{for}\ k=2^t\\
{\displaystyle{k_{\rm nn}(k)=4-\frac{3}{k}} } & \ \hbox{for}\
k=2^m+1 (m=2,3\cdots t).\end{array} \right.
\end{equation}
For those nodes with degree 3, it is easily to obtain
\begin{eqnarray}\label{knn03}
k_{\rm nn}(3)&=&\frac{2\cdot(2^t)^2+\sum_{\tau=1}^{\tau=t-1} \big
[L_v(\tau) k(\tau,t)\left[k(\tau,t)-1\right)\big]}{3 L_v(t)}+1
  \nonumber\\
  &=&\frac{4}{3}\,t+\frac{5}{3}-\frac{4}{3}\cdot\frac{1}{2^t},
\end{eqnarray}
where $k(\tau,t)$ is the degree of a node at time $t$ that was born
at step $\tau$. Thus $k_{\rm nn}(3)$ grows linearly with time for
large $t$. Eqs.~(\ref{knn02}) and (\ref{knn03}) show the network is
disassortative.

Degree correlation can also be described by a Pearson correlation
coefficient $r$ of degrees at either end of a link. It is defined as
\cite{Newman02,Newman03c,DoMa05,RaDoPa04}
\begin{equation}\label{Pearson}
r={\langle k\rangle\langle k^2 k_{\rm nn}(k)\rangle -
    \langle k^2\rangle^2 \over
    \langle k\rangle \langle k^3\rangle - \langle k^2\rangle^2}.
\end{equation}
We can easily see that for $t>1$, $r$ of $F_t$ is always negative,
indicating disassortativity.

\subsection{Average path length}

We represent all the shortest path lengths of $F_t$ as a matrix in
which the entry $d_{ij}$ is the shortest path from node $i$ to $j$.
A measure of the typical separation between two nodes in $F_t$ is
given by the average path length (APL) $\bar{d}_{t}$, also known as
characteristic path length~\cite{CoRoTrVi07}, defined as the mean of
geodesic lengths over all couples of nodes. APL is relevant in many
fields regarding real-life networks and has received much
attention~\cite{FrFrHo04,HoSiFrFrSu05,DoMeOl06,ZhChZhFaGuZo08}. In
the Appendix, we have obtained exact analytic expression for
$\bar{d}_{t}$, which reads
\begin{align}\label{APL}
\bar{d}_{t} = \frac{(16\cdot2^t+21)16^t+(21t-27)8^t+75\cdot4^t
+119\cdot2^t-15}{21 (2+5\cdot4^t+2\cdot 16^{t})}.
\end{align}
For large $t$, $\bar{d}_{t}\rightarrow \frac{8}{21}\cdot2^t$. Note
that in the infinite $t$ limit, $N_t \sim 4^t$, so the APL scales as
$\bar{d}_{t} \sim N_t^{1/2}$, which indicates that the network is
not a small world.

Thus, we have shown that $\bar{d}_{t}$ has the power-law scaling
behavior of the number of nodes $N_{t}$, which is similar to that of
two-dimensional regular lattice~\cite{Ne00}. This phenomenon is not
hard to understand. Let us look at the scheme of the network growth.
Each next step in the growth of $F_t$ doubles the APL between a
fixed pair of nodes (except those small number of pairs directly
connected by a noniterated link), while the total number of nodes
increases four-fold (asymptotically, in the infinite limit of $t$),
see Eq.~(\ref{Nt1}). Thus the APL $\bar{d}_{t}$ of $F_t$ grows as a
square power of the node number in the network.

\section{SIR model on the network}

As discussed in previous section, the network exhibits many
interesting properties, i.e., it is simultaneously scale-free,
highly clustered, ``large-world", fractal and disassortative, which
is not observed in uncorrelated networks with purely random wiring.
Therefore, it is worthwhile to investigate the processes taking
place upon the model to find the different impact on dynamic
precesses compared with uncorrelated networks. In what follows we
will study the SIR model of epidemics, which is one of the first
issues to be explored in the literature on complex
networks~\cite{PaVe01a,PaVe01b,MoPaVe02}.

In the standard SIR model~\cite{He00}, each node of the network
represents an individual and each link is the connection along which
the individuals interact and the epidemic can be transmitted. This
model describes diseases resulting in the immunization or death of
infected individuals, and assumes that each individual can be in one
of three possible states, namely, susceptible, infected, and
removed. The disease transmission on the network is described in an
effective way: At each time step, each susceptible node is infected
with probability $\lambda$, if it is connected to one or more
infected nodes; at the same time, each infected individual becomes
removed with probability 1, therefore it can not catch the infection
again.

The SIR model is equivalent to a bond percolation problem with bond
occupation probability $\lambda$~\cite{Gr82,Ne02}. Moreover, the
size of the outbreak is just the size of the giant component. In our
case the percolation problem can be solved using the real-space
renormalization group technique~\cite{MiEk7576,Ka76,Do03,RoAv07},
giving exact solution for the interesting quantity of epidemic
threshold. Let us describe the procedure in application to the
considered network. Assuming that the network growth stops at a time
step $t\rightarrow \infty$, when the network is spoiled in the
following way: for a link present in the undamaged network, with the
probability $\lambda$ we retain it in the damaged network. Then we
invert the transformation in Fig. 2 and define $n=t- \tau$ for this
inverted transformation, which is actually a decimation
procedure~\cite{Do03}. Further, we introduce the probability
$\lambda_n$ that if two nodes are connected in the undamaged network
at $\tau =t-n$, then at the $n$th step of the decimation for the
damaged network, there exists a path between these vertices. Here,
$\lambda_0= \lambda$. We can easily obtain the following recursion
relation for $\lambda_n$
\begin{eqnarray}\label{lambda01}
\lambda_{n+1}&=&\lambda_n^5+5\lambda_n^4(1-\lambda_n)+8\lambda_n^3(1-\lambda_n)^2+2\lambda_n^2(1-\lambda_n)^3
  \nonumber\\
  &=& 2\lambda_n^5-5\lambda_n^4+2\lambda_n^3+2\lambda_n^2.
\end{eqnarray}
Equation~(\ref{lambda01}) has an unstable fixed point at
$\lambda_c=\frac{1}{2}$, and two stable fixed points at $\lambda=0$
and $\lambda=1$.

Thus, for the SIR model the epidemic prevalence undergoes a phase
transition at a finite threshold $\lambda_c$ of the transmission
probability. If infection rate $\lambda >\lambda_c$, the disease
spreads and infects a finite fraction of the population. On the
other hand, when $\lambda <\lambda_c$, the total number of infected
individuals is infinitesimally small in the limit of very large
populations. The existence of an epidemic threshold in the present
network is compared to the result for uncorrelated scale-free
networks, where arbitrarily small infection rate $\lambda$ shows
finite prevalence~\cite{PaVe01a,PaVe01b,MoPaVe02}.

Why uncorrelated scale-free networks are prone to epidemics
spreading, while the present fractal disassortative scale-free
network can suppress disease propagation? This may be explained as
follows. In non-fractal uncorrelated scale-free topologies, the hubs
are connected and form a central compact core, such that the
infection of a few of the largest hubs has catastrophic consequences
for the network. For the fractal network, it is disassortative and
self-similar, which do not allow the presence of direct connections
between hubs, i.e., hubs are more dispersed (see Fig.~\ref{fig2}).
Thus, the mixture of fractal and  disassortative  properties
significantly provide protection against disease spreading. This
could provide insight into explaining why some real-life networks
have evolved into a fractal and disassortative
architecture~\cite{SoHaMa05,SoHaMa06}.

\section{Conclusions and discussion}

To conclude, we have investigated a class of deterministic graphs
from the viewpoint of complex networks. The deterministic
self-similar construction allow us to derive analytic exact
expressions for the relevant features. We have shown that the graphs
simultaneously exhibit many interesting structural characteristics:
power-law degree distribution, large clustering coefficient,
`large-world' phenomenon, fractal similar structure, negative degree
correlations. The simultaneous existence of scale-free, high
clustering, and `large-world' behaviors is compared with previous
network models.

Moreover, we have studied the SIR model in the graph under
consideration. We have presented the presence of a finite epidemic
threshold in the finite network size limit, showing that being prone
to disease spreading is not an intrinsic property of scale-free
networks. The ability of suppressing epidemic spreading may be owing
to its inherent topologies. Thus, our research may be helpful for
designing real networks resistent to epidemic outbreaks, and for the
better understanding of the role that network structure plays in the
spread of disease.

Although we have studied only a particular network corresponding to
$q=1$, in a similar way, one can easily investigate other networks
(i.e., $q\geq2$ cases) with various values of $\gamma$ and $d_B$,
and their general properties such as `large-world' behavior, high
clustering coefficient, and disassortative phenomenon are similar.
Analogously, one can also analyze the SIR model on top of these
networks. There is an existence of a different finite epidemic
thresholds for all the cases, which depend on the parameter $q$. We
speculate that the fractal property of the graphs determines the
presence of the threshold of disease transmission for SIR model that
can be mapped to a bond percolation as in the $d_B=\frac{\ln
(4q)}{\ln2}$ dimensional regular lattices~\cite{Ne00}. In the end,
we should mention that since most real-world networks are
stochastic, it would be interesting to construct random network
models displaying similar structural features as the deterministic
model studied in the present work.

\section*{Acknowledgment}

This research was supported by the National Basic Research Program
of China under grant No. 2007CB310806, the National Natural Science
Foundation of China under Grant Nos. 60496327, 60573183, 90612007,
60773123, and 60704044, the Postdoctoral Science Foundation of China
under Grant No. 20060400162, the Program for New Century Excellent
Talents in University of China (NCET-06-0376), and the Huawei
Foundation of Science and Technology (YJCB2007031IN).

\appendix

\section*{Appendix A: Derivation of the average path length}

\begin{figure}
\centering
\includegraphics[width=0.35\textwidth]{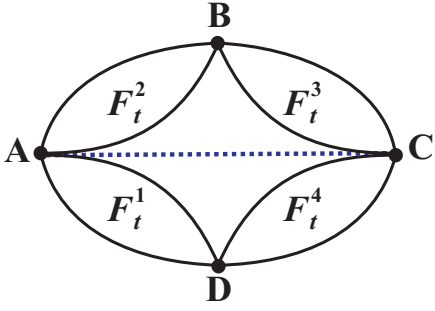}
\caption[kurzform] {(Color online) Second construction method of the
network that highlights self-similarity: The graph after $t+1$
    construction steps, $F_{t+1}$, is composed of four copies of
    $F_t$ denoted as
    $F_t^{\alpha}$ $(\alpha=1,2,3,4)$, which are
    connected to one another as above. The blue dashed link is noniterated
edge.}\label {apfig2}
\end{figure}

By definition, the APL for $F_t$ is defined as follows
\begin{equation}\label{eq:app4}
  \bar{d}_{t}  = \frac{D_t}{N_t(N_t-1)/2}\,,
\end{equation}
where
\begin{equation}\label{eq:app5}
  D_t = \sum_{i \in F_{t}, j \in F_{t}, i\neq j} d_{ij}
\end{equation}
denotes the sum of the chemical distances between two nodes over all
pairs, and $d_{ij}$ is the chemical distance between nodes $i$ and
$j$. The network has a self-similar structure allowing one to
calculate $\bar{d}_{t}$ analytically.  As shown in
Fig.~\ref{apfig2}, the lattice $F_{t+1}$ may be obtained by the
juxtaposition of four copies of $F_t$, which are labeled as
$L_{t}^{\alpha}$, $\alpha=1,2,3,4$. Then we can write the sum
$D_{t+1}$ as
\begin{equation}\label{eq:app6}
  D_{t+1} = 4\,D_t + \Delta_t\,,
\end{equation}
where $\Delta_t$ is the sum over all shortest paths whose endpoints
are not in the same $F_{t}$ branch. The solution of
Eq.~\eqref{eq:app6} is
\begin{equation}\label{eq:app8}
  D_t = 4^{t-1} D_1 + \sum_{x=1}^{t-1} 4^{t-x-1} \Delta_x\,.
\end{equation}
The paths that contribute to $\Delta_t$ must all go through at least
one of the four edge nodes ($\textbf{\emph{A}}$,
$\textbf{\emph{B}}$, $\textbf{\emph{C}}$, $\textbf{\emph{D}}$, see
Fig.~\ref{apfig2}) at which the different $F_t$ branches are
connected. The analytical expression for $\Delta_t$, called the
crossing paths, is found below.

Denote $\Delta_t^{\alpha,\beta}$ as the sum of all shortest paths
with endpoints in $F_t^{\alpha}$ and $F_t^{\beta}$. If
$F_t^{\alpha}$ and $F_t^{\beta}$ meet at an edge node,
$\Delta_t^{\alpha,\beta}$ rules out the paths where either endpoint
is that shared edge node.  If $F_t^{\alpha}$ and $F_t^{\beta}$ do
not meet, $\Delta_t^{\alpha,\beta}$ excludes the paths where either
endpoint is any edge node.  Then the total sum $\Delta_t$ is
\begin{align}
\Delta_t =& \,\Delta_t^{1,2} + \Delta_t^{1,3} + \Delta_t^{1,4}+
\Delta_t^{2,3} + \Delta_t^{2,4}+\Delta_n^{3,4}-2^{t+1}-1,
\label{eq:app10}
\end{align}
The last two terms at the end compensate for the overcounting of
certain paths: the shortest path between $\textbf{\emph{B}}$ and
$\textbf{\emph{D}}$, with length $2^{t+1}$, is included in
$\Delta_t^{1,2}$ and $\Delta_t^{3,4}$; the shortest path between
$\textbf{\emph{A}}$ and $\textbf{\emph{C}}$, with unit length 1, is
included in both $\Delta_t^{1,4}$ and $\Delta_t^{2,3}$.

By symmetry, $\Delta_n^{1,2} = \Delta_n^{3,4}$, $\Delta_t^{1,3} =
\Delta_t^{2,4}$ and $\Delta_t^{1,4} = \Delta_t^{2,3}$, so that
\begin{equation}\label{eq:app11}
\Delta_t = 2 \Delta_t^{1,2} + 2\Delta_t^{1,3} + 2\Delta_t^{1,4} -
2^{t+1} - 1,\,
\end{equation}
where $\Delta_t^{1,2}$ is given by the sum
\begin{align}
  \Delta_t^{1,2} &= \sum_{\substack{i \in F_t^{1},\,\,j\in
      F_t^{2}\\ i,j \ne A}} d_{ij}\nonumber\\
  &= \sum_{\substack{i \in F_t^{1},\,\,j\in
      F_t^{2}\\ i,j \ne A}} (d_{iA} + d_{Aj}) \nonumber\\
  &= (N_t-1)\sum_{i \in F_t^{1}} d_{iA} + (N_t-1) \sum_{j
    \in F_t^{2}} d_{Aj} \nonumber\\
  &= 2(N_t-1)\sum_{i \in F_t^{1}} d_{iA}\,,
\label{eq:app12}
\end{align}
where $\sum_{i \in F_t^{1}} d_{iA} = \sum_{j \in F_t^{2}} d_{Aj}$
has been used.  To find $\sum_{i \in
  F_t^{1}} d_{iA}$, we examine the structure of the
graph at the $t$th level.  In $F_t^{1}$, there are $\nu_t(m)$ points
with $d_{iA} = m$, where $1 \le m \le 2^t$, and $\nu_t(m)$ can be
written recursively as
\begin{equation}\label{eq:app13}
\nu_t(m) = \begin{cases} 2^t & \text{if $m$ is odd}\,,\\
\nu_{t-1}(\frac{m}{2}) & \text{if $m$ is even}\,.\end{cases}
\end{equation}
We can write $\sum_{i \in
  F_t^{1}} d_{iA}$ in terms of $\nu_t(m)$ as
\begin{equation}\label{eq:app14}
b_t \equiv \sum_{i \in F_t^{1}} d_{iA} = \sum_{m=1}^{2^t} m\cdot
\nu_t(m)\,.
\end{equation}
Eqs.~\eqref{eq:app13} and \eqref{eq:app14} relate $b_t$ and
$b_{t-1}$, which allow one to resolve $b_t$ by induction as follow:
\begin{eqnarray}
b_t &=& \sum_{k=1}^{2^{t-1}} (2k-1)\cdot2^t + \sum_{k=1}^{2^{t-1}}
2k\cdot \nu_{t-1}(k)\nonumber\\
&=& 2^{3t-2} + 2\,b_{t-1}=\frac{1}{3}\,2^{t} (2 + 4^{t})\,,
\label{eq:app15}
\end{eqnarray}
where $b_1 = \nu_1(1) + 2\nu_1(2)= 4$  has been used.  Substituting
Eq.~(\ref{eq:app15}) and $N_t = \frac{2}{3}(2+4^t) $ into
Eq.~(\ref{eq:app12}), we obtain
\begin{equation}\label{eq:app16}
\Delta_t^{1,2} =
\frac{1}{9}\,2^{t+1}\left(2^{1+2t}+1\right)(4^{t}+2)\,.
\end{equation}

Continue analogously,
\begin{align}
\Delta_t^{1,4} =& \sum_{\substack{i \in F_t^{1},\,i\ne A, D\\
j\in F_t^{4},\,j\ne C, D}} d_{ij}\nonumber\\
=& \sum_{\substack{i \in F_t^{1},\,i\ne A\\
j\in F_t^{4},\,j\ne C\\ d_{iA}+d_{jC} < 2^t}} (d_{iA}
   + d_{jC})+\sum_{\substack{i \in F_t^{1},\,i\ne A\\
j\in F_t^{4},\,j\ne C\\
d_{iA}+d_{jC} < 2^t}} d_{AC}    \nonumber\\
&+\sum_{\substack{i \in F_t^{1},\,i\ne A\\
j\in F_t^{4},\,j\ne C\\
d_{iC}+d_{jC} = 2^t}} 2^t+\sum_{\substack{i \in F_t^{1},\,i\ne A\\
j\in F_t^{4},\,j\ne C\\ d_{iC}+d_{jC} > 2^t}} (d_{iD}
   + d_{jD})\nonumber\\
&+\sum_{\substack{i \in F_t^{1}}}d_{iC}+\sum_{\substack{j \in
F_t^{4}}}d_{jA}-2\cdot 2^t\ - 1, \label{eq:app17}
\end{align}
The first term equals the fourth one and is denoted by $g_t$. The
second and third terms are denoted by $y_t$ and $h_t$, respectively.
The fifth and sixth terms are equal to each other, both of which
equal to $b_t+N_t-1$. The last two terms at the end compensate for
the redundancy or overcounting of certain paths: the shortest paths
of $d_{CD}$ in $\sum_{i \in F_t^{1}}d_{iC}$ and $d_{AD}$ in $\sum_{j
\in F_t^{4}}d_{jA}$ are redundant, both of which are $2^t$; the
shortest path between $d_{AC}$ with unit length 1 is included in
both $\sum_{i \in F_t^{1}}d_{iC}$ and $\sum_{j \in F_t^{4}}d_{jA}$.
 Therefore $\Delta_t^{1,4} = 2g_t + h_t+y_t+2b_t-2^{t+1}+ 2 N_t -3$. One can compute the quantity $g_t$ as
\begin{align}
  g_t =& \sum_{m=1}^{2^t-2}\:\sum_{m^\prime = 1}^{2^t-1-m} \nu_t(m)\nu_t(m^\prime)(m+m^\prime)\nonumber\\
  =& \sum_{k=1}^{2^{t-1}-2}\:\sum_{k^\prime = 1}^{2^{t-1}-1-k} \nu_{t-1}(k)\nu_{t-1}(k^\prime)(2k+2k^\prime)\nonumber\\
  &+ \sum_{k=1}^{2^{t-1}-1}\:\sum_{k^\prime = 1}^{2^{t-1}-k} \nu_{t-1}(k)2^t (2k+2k^\prime-1)\nonumber\\
  &+ \sum_{k=1}^{2^{t-1}-1}\:\sum_{k^\prime = 1}^{2^{t-1}-k} 2^t \nu_{t-1}(k^\prime) (2k-1+2k^\prime)\nonumber\\
  &+ \sum_{k=1}^{2^{t-1}-1}\:\sum_{k^\prime = 1}^{2^{t-1}-k} 2^{2t}
  (2k-1+2k^\prime-1)\,,
\label{eq:app18}
\end{align}
where the fourth term can be summed directly, yielding
\begin{align}
\frac{1}{3}\,2^{-2+3 t}(-2+2^t)(-1+2^t). \label{eq:app19}
\end{align}
In Eq.~\eqref{eq:app18}, the second and third terms are equal to
each other and can be simplified by first summing over $k^\prime$,
yielding
\begin{align}
2^t\,\sum_{k=1}^{2^{t-1}-1} \nu_{t-1}(k) \left(4^{t} - k^2\right)\,.
\label{eq:app20}
\end{align}
For use in Eq.~\eqref{eq:app20}, $\sum_{k=1}^{2^{t-1}-1}
\nu_{t-1}(k) = N_{t-1}-2$. On the other hand, it is easy to derive
the following recursive relation
\begin{align}
  \sum_{k=1}^{2^{t-1}} k^2 \nu_{t-1}(k)=
  \sum_{z=1}^{2^{t-2}} (2z-1)^2 2^{t-1} + \sum_{z=1}^{2^{t-2}}(2z)^2
  \nu_{t-2}(k),
 \end{align}
 using which we have
\begin{align}
  \sum_{k=1}^{2^{t-1}-1} k^2 \nu_{t-1}(k) &= \sum_{k=1}^{2^{t-1}} k^2
  \nu_{t-1}(k) - 2^{2t-2}\nonumber\\ &=
  \frac{1}{9}2^{2t-3}(4^t-3t+17)-2^{2t-2}.
 \label{eq:app21}
\end{align}
With these results, Eq.~\eqref{eq:app20} becomes
\begin{align}
\frac{1}{9}\,8^{-1+t}\left(-11 + 2^{1+2t} + 3t\right)\,.
\label{eq:app23}
\end{align}
With Eqs.~\eqref{eq:app19} and \eqref{eq:app23},
Eq.~\eqref{eq:app18} becomes
\begin{align}
g_t =& 2g_{t-1} +
\frac{2^t}{36}\left(5\cdot16^t-5\cdot4^t-9\cdot8^t+3t\cdot4^t\right).
\label{eq:app24}
\end{align}
Considering the initial condition $g_1 = 0$, we can solve
Eq.~\eqref{eq:app24} inductively leading to
\begin{align}
g_t=\frac{1}{945} 2^{t}
\left(158+112\cdot16^{t}-270\cdot8^t+210t\cdot4^{t -1}\right)\,.
\label{eq:app25}
\end{align}
We now evaluate $h_t$ using a recursive method:
\begin{align}
h_t &= 2^{t} \sum_{m=1}^{2^t-1} \nu_t(m) \nu_t(2^t-m)\nonumber\\
&= 2^{t} \sum_{m=1}^{2^t-1} \nu^2_t(m)\nonumber\\
&= 2^{t} \left[\sum_{k=1}^{2^{t-1}} 4^t  + \sum_{k=1}^{2^{t-1}-1}
  \nu^2_{t-1}(k)\right]\nonumber\\
&= 2^{4t-1} + 2h_{t-1}\,, \label{eq:app26}
\end{align}
where we have used the the symmetry $\nu_t(m) = \nu_t(2^t-m)$. Since
$h_1 = 8$, Eq.~\eqref{eq:app26} is solved inductively:
\begin{equation}
h_t = 2^{t+2}(8^t-1)/7\,. \label{eq:app27}
\end{equation}
To find an expression for $\Delta_t^{1,4}$, now the only thing left
is to evaluate $y_t$, which can be calculated using the same way as
$g_t$. Replacing $m+m'$ in Eq.~(\ref{eq:app18}) with 1, one can get
\begin{align}
  y_t =& \sum_{m=1}^{2^t-2}\:\sum_{m^\prime = 1}^{2^t-1-m} \nu_t(m)\nu_t(m^\prime)\nonumber\\
  =& \sum_{k=1}^{2^{t-1}-2}\:\sum_{k^\prime = 1}^{2^{t-1}-1-k} \nu_{t-1}(k)\nu_{t-1}(k^\prime)\nonumber\\
  &+ \sum_{k=1}^{2^{t-1}-1}\:\sum_{k^\prime = 1}^{2^{t-1}-k} \nu_{t-1}(k)2^t \nonumber\\
  &+ \sum_{k=1}^{2^{t-1}-1}\:\sum_{k^\prime = 1}^{2^{t-1}-k} 2^t \nu_{t-1}(k^\prime)\nonumber\\
  &+ \sum_{k=1}^{2^{t-1}-1}\:\sum_{k^\prime = 1}^{2^{t-1}-k} 2^{2t}
  (2k-1+2k^\prime-1)\,.
\label{eq:app28}
\end{align}
Analogously to the computation of $g_t$, we can easily obtain
\begin{equation}
y_t =
\frac{2}{63}(-1+2^t)\left(7\cdot8^{t}-2\cdot4^{t}-16\cdot2^{t}-16\right)
\end{equation}
Combining previous equations and results, we get the final
expression for $\Delta_t^{1,4}$,
\begin{align}
\Delta_t^{1,4} &=\frac{1}{189} \Big (33-98\cdot 2^{t}+168\cdot 2^{2
t}-12\cdot 2^{3 t} -66 \cdot 2^{4 t}\nonumber\\ &\quad +56\cdot 2^{5
t}+ 108 \cdot 4^{2 t}+42t\cdot 2^{3 t} \Big) \,. \label{eq:app29}
\end{align}
We now begin to compute $\Delta_t^{1,3}$, which can be obtained from
$\Delta_t^{1,2}$ by regarding $A$ and $C$ as one single point, plus
$d_{AC}$ to the path length of all related pairs of nodes, so that
\begin{align}
\Delta_t^{1,3} &= \Delta_t^{1,2} + (N_t-2)^2 - 2\left[b_t + 2^t\left(N_t-1\right)\right]+2^{t+1}\nonumber\\
&= \frac{4}{9}\Big(1+2^t-2^{1+2 t}-2^{1+3 t}+2^{4t}+2^{5t}\Big) \,.
\label{eq:app30}
\end{align}
Substituting Eqs.~(\ref{eq:app16}), (\ref{eq:app29}) and
(\ref{eq:app30}) into Eq.~(\ref{eq:app11}), we obtain the final
expression for the crossing paths $\Delta_t$:
\begin{align}
 \Delta_t = \frac{1}{189} \Big[45 &-119\cdot 2^{t+1}+15\cdot 2^{3 t+2}+7\cdot2^{5 t+6}\nonumber\\
  &+63\cdot4^{2 t+1}+21t\cdot2^{3 t+2} \Big]\,. \label{eq:app31}
\end{align}
Substituting Eqs.~(\ref{eq:app31}) for $\Delta_x$ into
Eq.~(\ref{eq:app8}), and using $D_0 = 1$, we have
\begin{align}
D_t = \frac{1}{189} \Big[2^{4+5 t}&+21\cdot 2^{4 t}+21t\cdot 2^{3
t}-27\cdot 2^{3 t}\nonumber\\&+75\cdot 2^{2t} +119\cdot 2^t -15
\Big] \,.\label{eq:app32}
\end{align}
Inserting Eq.~\eqref{eq:app32} into Eq.~\eqref{eq:app4}, one can
obtain the analytical expression for $\bar{d}_t$ in Eq.~\eqref{APL}.


\end{document}